\documentclass[aps,prl,twocolumn,showpacs,floatfix,superscriptaddress,amsmath,amssymb]{revtex4}
\usepackage{graphicx}

\begin{document}

\title{Nanoscale superconducting properties of amorphous W-based deposits grown with focused-ion-beam}

\author{I. Guillam\'on}
\affiliation{Laboratorio de Bajas Temperaturas, Departamento de F\'isica de la Materia Condensada \\ Instituto de Ciencia de Materiales Nicol\'as Cabrera, Facultad de Ciencias \\ Universidad Aut\'onoma de Madrid, E-28049 Madrid, Spain}
\author{H. Suderow}
\affiliation{Laboratorio de Bajas Temperaturas, Departamento de F\'isica de la Materia Condensada \\ Instituto de Ciencia de Materiales Nicol\'as Cabrera, Facultad de Ciencias \\ Universidad Aut\'onoma de Madrid, E-28049 Madrid, Spain}
\author{S. Vieira}
\affiliation{Laboratorio de Bajas Temperaturas, Departamento de F\'isica de la Materia Condensada \\ Instituto de Ciencia de Materiales Nicol\'as Cabrera, Facultad de Ciencias \\ Universidad Aut\'onoma de Madrid, E-28049 Madrid, Spain}
\author{A. Fern\'andez-Pacheco}
\affiliation{Instituto de Nanociencia de Arag\'on, Universidad de Zaragoza, Zaragoza, 50009, Spain}
\affiliation{Instituto de Ciencia de Materiales de Arag\'on, Universidad de Zaragoza-CSIC, Facultad de Ciencias, Zaragoza, 50009, Spain}
\affiliation{Departamento de F\'isica de la Materia Condensada, Universidad de Zaragoza, 50009 Zaragoza, Spain}
\author{J. Ses\'e}
\affiliation{Instituto de Nanociencia de Arag\'on, Universidad de Zaragoza, Zaragoza, 50009, Spain}
\affiliation{Departamento de F\'isica de la Materia Condensada, Universidad de Zaragoza, 50009 Zaragoza, Spain}
\author{R. C\'ordoba}
\affiliation{Instituto de Nanociencia de Arag\'on, Universidad de Zaragoza, Zaragoza, 50009, Spain}
\affiliation{Departamento de F\'isica de la Materia Condensada, Universidad de Zaragoza, 50009 Zaragoza, Spain}
\author{J.M. De Teresa}
\affiliation{Instituto de Ciencia de Materiales de Arag\'on, Universidad de Zaragoza-CSIC, Facultad de Ciencias, Zaragoza, 50009, Spain}
\affiliation{Departamento de F\'isica de la Materia Condensada, Universidad de Zaragoza, 50009 Zaragoza, Spain}
\author{M.R. Ibarra}
\affiliation{Instituto de Nanociencia de Arag\'on, Universidad de Zaragoza, Zaragoza, 50009, Spain}
\affiliation{Instituto de Ciencia de Materiales de Arag\'on, Universidad de Zaragoza-CSIC, Facultad de Ciencias, Zaragoza, 50009, Spain}
\affiliation{Departamento de F\'isica de la Materia Condensada, Universidad de Zaragoza, 50009 Zaragoza, Spain}

\begin{abstract}

We present very low temperature Scanning Tunneling Microscopy and Spectroscopy (STM/S) measurements in W-based amorphous superconducting nanodeposits grown using a metal-organic precursor and focused-ion-beam. The superconducting gap closely follows s-wave BCS theory, and STS images under magnetic fields show a hexagonal vortex lattice whose orientation is related to features observed in the topography through STM. Our results demonstrate that the superconducting properties at the surface of these deposits are very homogeneous, down to atomic scale. This, combined with the huge nanofabrication possibilities of the focused-ion-beam technique, paves the way to use focused-ion-beam to make superconducting circuitry of many different geometries.

\end{abstract}

 \date{\today}

\maketitle

Since the seminal work of Buckel \cite{Buckel56}, which found superconductivity at 7 K in fully amorphous films of Bi (Bi is not superconducting as a crystalline solid at ambient pressure), many materials have been found to deeply change their electrical properties when prepared in an amorphous form \cite{Johnson82}. The actual origin of superconductivity, and the relationship to the metal-insulator transition induced by strong disorder, has been subject of intense debate\cite{Lee85}. However, some of these systems have metallic transport and thermodynamic properties that better approach the free electron model than their crystalline analogs \cite{Johnson82}. Moreover, those which are superconducting very often behave as model s-wave BCS superconductors \cite{Johnson82}. This makes them particularly interesting candidates as a reference system where s-wave BCS phenomenology can be probed in newly developed techniques, in particular, in nanoscale tunneling probes of the superconducting state as STM/S. Many crystalline superconductors such as High T$_c$ cuprates, MgB$_2$, heavy fermions, transition metal di-chalcogenides, boron doped diamond, or nickel borocarbides, have been studied using high resolution STM/S down to atomic level and in a wide range of temperatures and magnetic fields \cite{Fischer07,Pan00,Gomes07,Nishida07,Sacepe06,Hess89,Rubio01,Suderow04,Suderow01}. However, an atomic level STM/S experiment over large surfaces in a reference s-wave BCS superconductor is still lacking. In this article we report a thorough STM/S study at atomic scale of amorphous W nanodeposits. We find that the nanodeposits remain fully stable under ambient conditions, showing excellent local tunneling properties. The nanodeposited amorphous system is chemically highly inert, as e.g. 2H-NbSe$_2$ or graphite, making this system the long sought paradigm for STM/S studies, with the additional important feature of being easily used for creating nanofabricated superconducting structures.

The W nanodeposits have been grown with a dual-beam system that integrates an electron and an ion column. A major application of this system is the controllable nanodeposition of materials, with widely extended use in nanotechnology and the semiconductor industry for the reparation of lithography masks and integrated circuits \cite{Matsui96}, the fabrication of three-dimensional nanostructures \cite{Puers01}, the creation of electronic nanodevices \cite{deMarco04}, the protection of sample surface for subsequent micromachining for cross-section analysis \cite{Gianuzzi05}, etc. In the case of ion-beam-induced-deposition (IBID), an ion beam focused to nanometric size (generally based on Ga$^+$) is swept through the region of interest in a sample, breaking precursor gas molecules previously introduced in the chamber and giving rise to a nanodeposit. The combined use of imaging techniques (secondary-electron or backscattered-electron images) and local growth techniques (IBID) allow precise and flexible deposition in the targeted place, with the lateral size and deposit thickness being controlled at a large extent. The IBID growth process can be considered a chemical vapour deposition process assisted by an ion beam. The precursor gas molecules containing the material to be deposited, flow from a nearby injector towards the substrate and become adsorbed to it. The mechanism proposed to explain the deposition is the collision cascade model where the ions impact on the substrate and transfer energy to surface atoms, which decompose the precursor gas molecules \cite{Ro94}. The role played by the secondary electrons reaching the sample surface has been also highlighted \cite{Lipp96}. When the precursor molecules are organic, a carbonaceous matrix is also present, which is relevant for the physical and chemical properties exhibited by the deposit. With the IBID technique, nanodeposits based on metallic materials such as Au, Pt, W and Cu have been carried out and used in nanotechnological applications \cite{Gianuzzi05,Ro94,Lipp96,Ratta93,Bannerjee93,Lin03}. Amongst IBID materials, W nanodeposits have been recently found to show superconducting properties \cite{Sadki04,Sadki05,Luxmoore07,Spoddig07}.

\begin{figure}
\includegraphics[angle=270,width=10cm,clip]{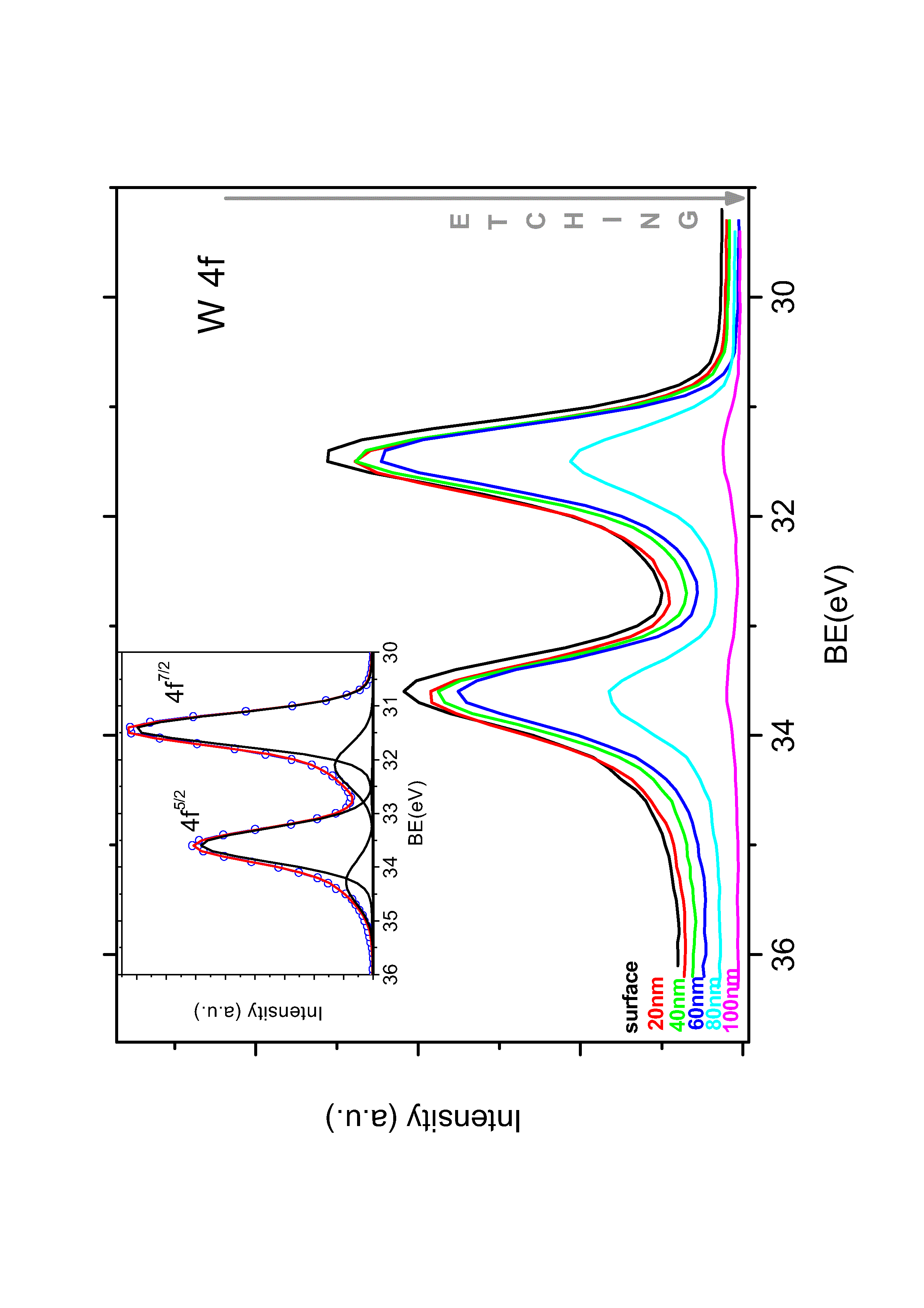}
\caption{Depth profile spectra of the W 4f core level. Successive spectra are obtained after a 20 nm argon etching process. A typical fit is shown in the inset. The solid lines are the components in which the spectrum is decomposed. The resulting fit is superimposed to the experimental data (open circles). The energies found for the peaks are associated to metallic W and WC (see text). The proportion between both species (in average W/WC=5.2(5)) remains approximately constant through all the deposit thickness. For the quantitative analysis of the XPS spectra, we have used pseudo-Voigt peak profiles with a 10\% to 20\% Lorentzian contribution, subtracting a Shirley - type background. \label{Fig2}}
\end{figure}

The STM/S measurements were performed on a 200 nm-thick film of a W deposit with a lateral size 30 $\mu$m $\times$ 30 $\mu$m grown at the center of a conducting Au layer previously deposited on a Si substrate. The deposit has been made in a similar way to previous reports \cite{Sadki04,Sadki05,Luxmoore07}, and characterized by means of energy dispersive X-Ray microanalysis (EDX), high resolution transmission electron microscopy (HRTEM), transport measurements and X-Ray photoelectron spectroscopy (XPS). The deposit is metallic due to the high content of W (around 40\% from EDX), as found previously in nanodeposits of close composition \cite{Sadki04,Sadki05,Luxmoore07}. An in-depth study of the chemical composition using XPS and an argon etching process shows that the deposit is very homogeneous, with average concentration W=40$\pm$7\%, C=43$\pm$4\%, Ga=10$\pm$3\% and O=7$\pm$2\%. In Fig.\ \ref{Fig2} we can see the W 4f spectrum as a function of the sputtering depth. In the inset, a typical fit is shown. The main peak for the W4f$^{7/2}$ appears at 31.4 eV, associated to metallic W\cite{Luthin00,Xu07}. A minor peak at 32.1 eV, corresponding to tungsten carbide (WC)\cite{Luthin00,Xu07} is also present. A separation of 2.1eV between the 4f$^{7/2}$ and the 4$^{f5/2}$ components is found in all cases. The ratio between both species through the thickness is W/WC=5.2$\pm$0.5. The C 1s spectrum (not shown here) is significantly broad, and can be de-convoluted in three peaks. One at 283.1eV, due to WC \cite{Diaz96,Haerle02}, and the others at 284.3 and 285.2eV, which correspond to amorphous carbon, sp$^2$ and sp$^3$ atoms, respectively\cite{Diaz96,Haerle02}. The proportion between both types of carbon (sp$^2$/sp$^3$) corresponds to 3.45$\pm$0.15 through the profile.

The sample used here was stored for two days under ambient conditions, without any additional surface treatment and then mounted in a low temperature STM thermally anchored to the mixing chamber of a dilution refrigerator. The magnetic field was applied perpendicular to the sample surface. Our set-up, described in previous work, has a very high mechanical stability and we can change in-situ the scanning window by moving the sample with respect to the tip \cite{Suderow04,Rodrigo04b,Crespo06a,Guillamon08}. The gold tip was mechanically shaped and then carefully positioned at ambient temperature on top of the nanodeposit. At low temperatures, when approaching the tip to tunneling distance, we immediately found that the tunneling conditions of the nanodeposit were outstanding. The work function was always high valued (several eV), and the images and spectra were fully reproducible and independent of the tunneling current (or distance between tip and sample).

\begin{figure}
\includegraphics[width=8cm]{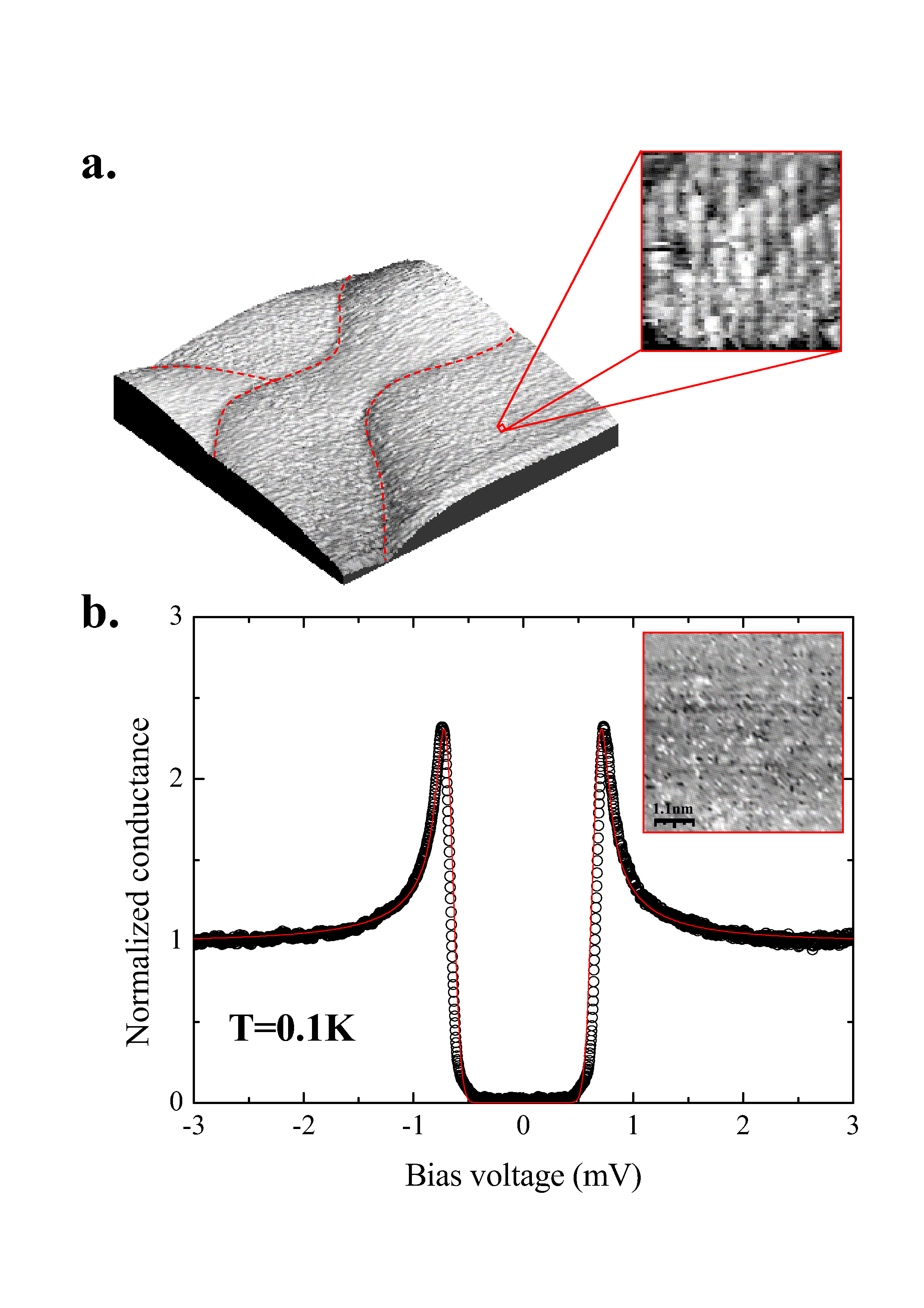}
\caption{A typical topographic STM image is shown in a (size 1 $\mu$m x 1 $\mu$m). Red dashed lines mark the small linear depressions found over the whole surface. Maximum contrast corresponds to height differences of the order of 10 nm, i.e. 5$\%$ of the film width. The linear depressions separate atomically flat surfaces. In the inset we show an atomic size topographic STM image (size 10 nm x 10 nm). In b. we show the tunneling conductance versus bias voltage typically found over the whole image shown in a. Red line is a fit to BCS theory using $\Delta$=0.66 meV. The inset shows a STS image made in the same area as the inset of a. The contrast corresponds to changes of about 10\% in the tunneling spectra at the voltage position of the quasiparticle peaks.\label{Fig3}}
\end{figure}

\begin{figure}
\includegraphics[width=8cm]{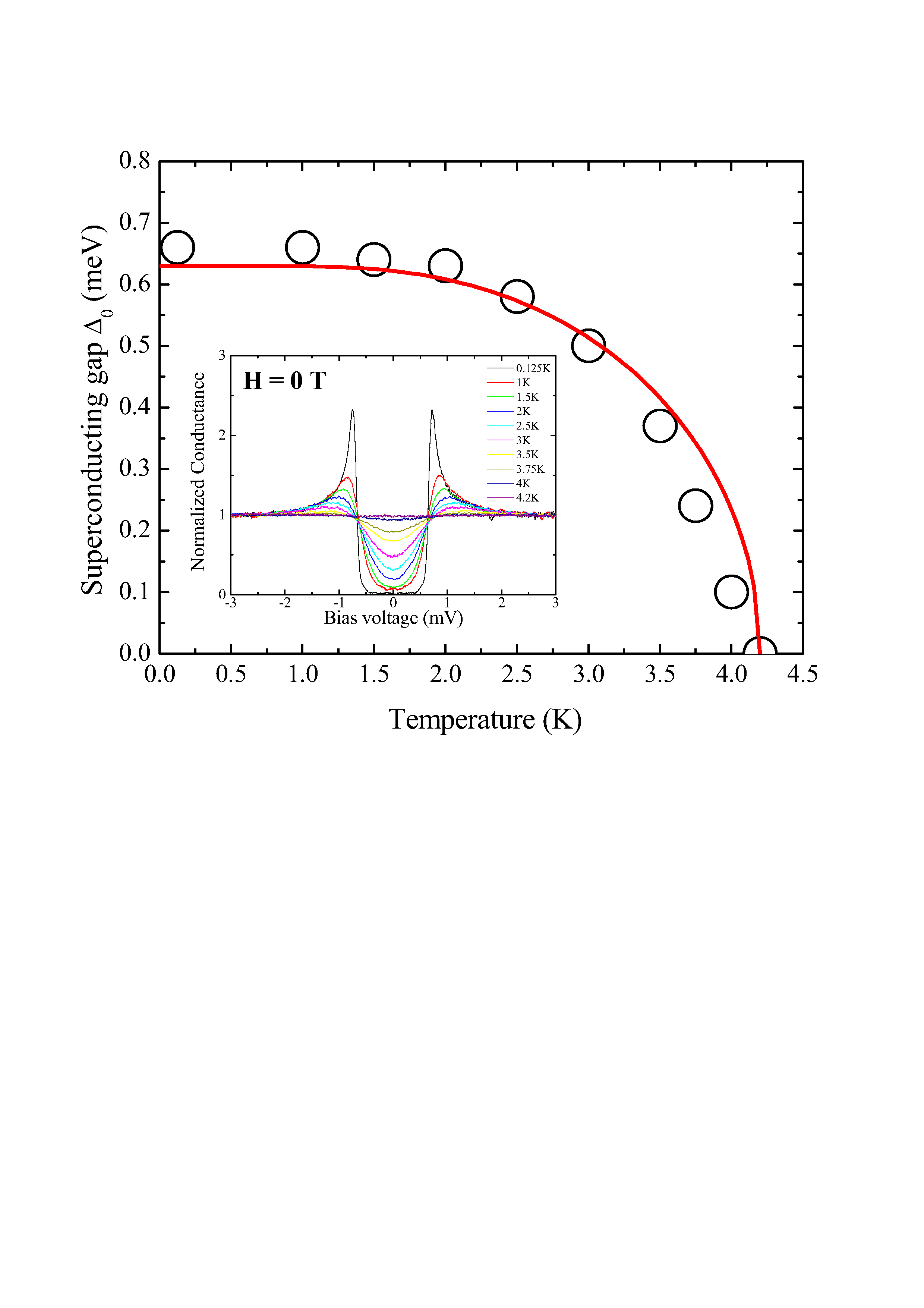}
\vskip -5cm
\caption{Temperature dependence of the superconducting gap obtained from fits to the tunneling spectra shown in the inset. Red line is BCS expression, taking $\Delta$=1.73 k$_B$T$_c$ and T$_c$=4.15 K.\label{Fig4}}
\end{figure}

Topographic images show that it is possible to scan large areas (Fig.\ref{Fig3} a), up to some $\mu$m$^2$, obtaining very uniform scanning profiles. Large regions with surface roughness on the nm range or below and a small inclination (below 5$^{\circ}$) appear surrounded by long linear depressions of at most 5-10 nm depth, i.e. between 2 and 5 \% of the total film thickness (red lines in Fig.\ref{Fig3} a). STM images in the nm range (inset of Fig.\ref{Fig3}a) show small groups of randomly distributed atomic size features. The tunneling conductance curve shown in Fig.\ref{Fig3}b (taken at 0.1 K) is obtained over whole surface areas, without any remarkable changes. The fit to BCS theory is excellent, and gives a value of $\Delta$ = 0.66 meV for the superconducting gap, which is slightly higher than the weak coupling BCS value $\Delta_0$ = 1.73 k$_B$T$_c$ = 0.63 meV (T$_c$=4.15 K). The voltage dependence of the conductance as a function of temperature gives the temperature dependence of the superconducting gap, shown in Fig.\ref{Fig4}, which follows well expectations from BCS theory (solid line in Fig.\ref{Fig4}).

The inset of Fig.\ref{Fig3}b shows a STS image constructed from the conductance at the quasiparticle peak in the same region in which the topographic STM image shown in the inset of Fig.\ref{Fig3}a was taken. The contrast between black and white corresponds to changes of about 10\% of the value of the conductance at the quasiparticle peaks. Corresponding fits to BCS theory show associated changes in $\Delta$ below 2\%. Clearly, the atomic size features observed in the topography (STM, inset of Fig.\ref{Fig3}a) are not seen in the spectroscopic imaging (STS, inset of Fig.\ref{Fig3}b). Similar results are observed when comparing STM and STS images at $\mu$m length scale. Therefore, the superconducting density of states N(E,\textbf{r}) can be considered to be homogeneous, in particular if we compare our results to those found in atomic size STS experiments in other kinds of superconducting materials, e.g. the High-T$_c$ cuprate superconductors or 2H-NbSe$_2$. In the cuprate superconductors, periodic modulations of  N(E,\textbf{r}) due to surface scattering have provided a great deal of new insight into important properties of the superconducting state, as e.g. the anisotropy of the superconducting gap \cite{Fischer07,Pan00,Gomes07}. In 2H-NbSe$_2$, we could recently demonstrate that the two-band superconducting properties of this compound result in small modulations of N(E,\textbf{r}) that follow lattice periodicity \cite{Guillamon08}. The absence of any of these effects here demonstrates that the nanodeposit is the first fully isotropic s-wave superconductor that can be studied in detail down to atomic level.

On the other hand, note that the compositional analysis performed with XPS along the full sample thickness (see Fig.\ref{Fig2}) demonstrates that the composition is homogeneous as a function of the depth, up to the surface. This is in clear correspondence with the dramatic uniformity of the tunneling characteristics over the whole surface, and demonstrates that the superconducting properties of these deposits have a remarkable overall homogeneity, as may be expected for an amorphous system.

\begin{figure}
\includegraphics[width=8cm]{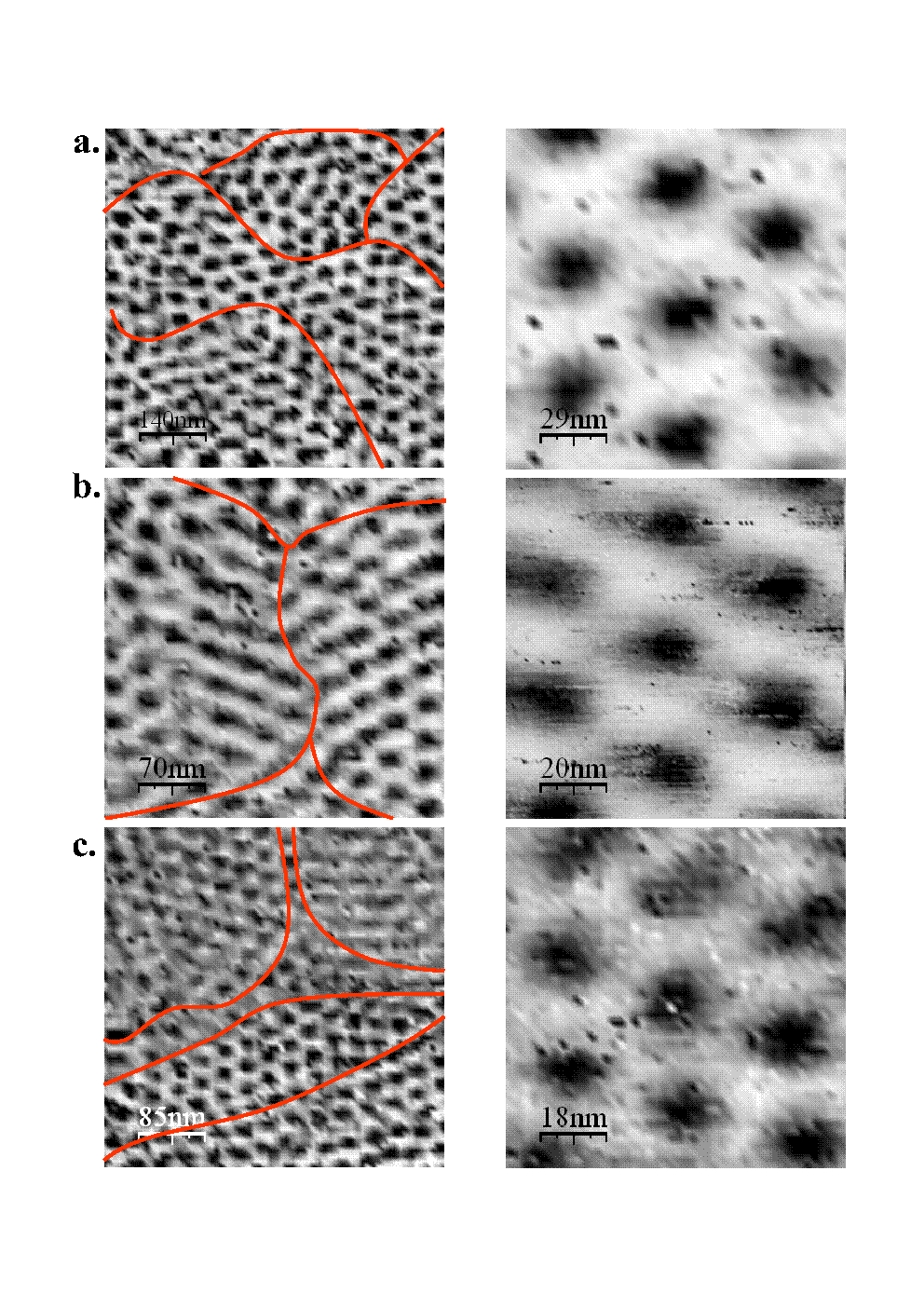}
\caption{Vortices found at different magnetic fields (a. at 1 T, b. at 2 T and c. at 3 T) obtained from STS images constructed from the Fermi level conductance. Left panels show large STS images, and right panels show a single hexagon. The topographic STM images corresponding to the left panels have the small linear surface depressions characteristic of these samples (as discussed in the text and shown in Fig.\protect\ref{Fig3}a), along the red lines. The STS images of the right panels are made far from surface depressions, and the corresponding topographic STM images are absolutely flat and featureless, with surface roughness below about 0.5 nm.\label{Fig5}}
\end{figure}

When we apply a magnetic field, we observe that, below about 0.2 T, no clear hexagonal vortex lattice is formed in the sample. Vortices appear bunched close to the linear surface depressions observed in the topography (Fig.\ref{Fig3}a). Those lying far from the linear depressions, at the smooth and very flat regions surrounded by them, are observed to easily change their position, especially some minutes after changing the field. Above about 0.3 T, however, a clear and very stable vortex lattice is formed. In Fig.\ref{Fig5} we show some representative examples of STS vortex images obtained from the zero bias conductance changes as a function of the position. The vortex lattice is observed over the full scanning window, and has no long range order, although, in between the linear depressions in the associated topographic images (marked by red lines in the Fig.\ref{Fig5}), a clear hexagonal Abrikosov lattice is often observed. The right panels of Fig.\ref{Fig5} are single hexagons obtained far from the linear surface depressions. The intervortex distance $d$ within these hexagons follows well conventional expectations for a superconductor with a hexagonal vortex lattice ($d=(1.25)^{1/4}(\frac{\Phi_0}{H})^{1/2}$, where $\Phi_0$ is the flux quantum\cite{T96}).

Vortices have a strong tendency to form rows parallel to the linear surface depressions. Correspondingly, the orientation of the hexagons far from the linear depressions slightly changes in different areas. In spite of being very small and smooth changes in the topography of the sample, the surface depressions are clearly the only source of distortion of the Abrikosov lattice in these samples. Indeed, the pinning properties of transition-metal amorphous systems have been studied thoroughly, and the weak bulk pinning ensures that surface features play a dominant role in the potential landscape that determines the vortex positions \cite{Kes83}.

Note that vortices appear sometimes as elongated figures in the images, especially close to the linear depressions. The vortex density in our experiments is high and $d$ is much smaller than the film thickness. So vortex lines can curve in order to hit the surface at a right angle, as the curved length, of the order of $d$, is small \cite{Brandt93}. This effect can easily cause elongated vortex forms in an STS experiment, as e.g. in Fig.\ref{Fig5}b. Note that this is in sharp contrast to previous vortex imaging experiments in thin films, performed at much smaller magnetic fields using techniques sensitive to the local magnetic field \cite{Plourde02,Bending99}. There, vortex tilting effects at the surface significantly increase the vortex length and are no longer favorable.

Let us put forward that these W IBID nanodeposits can find outstanding applications in nanotechnology. Here we have shown that these deposits remain clean after exposure to ambient conditions, have excellent and homogeneous s-wave BCS superconducting properties going up to the surface, and that the vortex configuration is given by the potential landscape formed by small height changes at the surface. The flexible and controllable growth of this superconductor with the desired design onto a targeted place by means of a fine nanometric ion beam in combination with the extraordinary stable surface pave the way for new and unexplored devices based on nanoscopic superconductivity. For example, nanostructured islands that may serve in quantum computation problems, or in vortex confinement, stable micro and nano SQUID's, hybrid superconductor-ferromagnet systems, or stable connecting arrangements to measure isolated molecular systems.

We are indebted to A. Mel'nikov, A.I. Buzdin, F. Guinea, J.G. Rodrigo, V. Crespo and J.P. Brison for discussions and J. Arbiol for assistance in the HRTEM experiments. The Laboratorio de Bajas Temperaturas is associated to the ICMM of the CSIC. This work was supported by the Spanish MEC (Consolider Ingenio 2010, MAT and FIS programs), by the Comunidad de Madrid through program "Science and Technology in the Millikelvin", and by NES and ECOM programs of the ESF.


\end{document}